\documentclass[aps,prl,twocolumn,showpacs]{revtex4}
\usepackage{graphicx,amssymb}

\begin{document}

\title{Observation of excited states in a p-type GaAs quantum dot}

\author{Y. Komijani$^{1}$, M. Csontos$^{1}$, T. Ihn$^{1}$, K. Ensslin$^{1}$, D. Reuter$^{2}$ and A. D. Wieck$^{2}$}

\affiliation{$^{1}$Solid State Physics Laboratory, ETH Z\"urich - 8093 Z\"urich, Switzerland\\
$^{2}$Angewandte Festk\"orperphysik, Ruhr-Universit\"at Bochum - 44780 Bochum, Germany}

\date{\today}

\begin{abstract}

A quantum dot fabricated by scanning probe oxidation lithography on a p-type, C-doped GaAs/AlGaAs heterostructure is investigated by low temperature electrical conductance measurements. Clear Coulomb blockade oscillations are observed and analyzed in terms of sequential tunneling through the single-particle levels of the dot at $T_{\rm hole}$ = 185 mK. The charging energies as large as $\sim$ 2 meV evaluated from Coulomb diamond measurements together with the well resolved single-hole excited state lines in the charge stability diagram indicate that the dot is operated with a small number of confined particles close to the ultimate single-hole regime.

\end{abstract}


\maketitle

\maketitle


Quantum dots implemented in GaAs heterostructures represent promising candidates for the experimental realization of quantum computation \cite{Loss98}, as well as various spintronic devices \cite{Wolf01}. However, research based on electronic transport through such small conducting islands has been, so far, almost exclusively focused on quantum dots defined on n-type GaAs heterostructures \cite{Kouwenhoven97}. The recently emerging interest in low-dimensional hole-doped systems arises primarily from the fact that spin-orbit as well as carrier-carrier Coulomb interaction ($E_{\rm int}$) effects are more pronounced in such systems compared to the more established n-doped systems. The main reason for this is that holes have a much higher effective mass than electrons, and thus a smaller Fermi energy $E_{\rm F}$. This enables the investigation of novel regimes with much higher interaction parameter $r_s=E_{\rm int}/E_{\rm F}$.

Stronger spin-orbit interactions in bulk two-dimensional hole-doped systems are expected to lead to significantly reduced spin relaxation times. On the other hand, it was also shown \cite{Schneider04} that spin relaxation of holes confined into quantum wells is much slower than in the bulk case, but still several orders of magnitude faster than electron spin relaxation. This was one of the main reasons why p-type systems received little attention in efforts to utilize the hole spins in quantum information technologies. However, as it was recently predicted \cite{Bulaev05}, further confinement of holes into quantum dots can significantly increase the relaxation time $T_{1}$ of hole spins, so that it can be comparable, or even larger than the one of the electron spins. In this paper we present the results of Coulomb blockade measurements in such a single-hole transistor defined on a p-type carbon doped GaAs heterostructure. The area occupied by the confined holes in the investigated device turns out to be much smaller than in previously reported p-type quantum dots \cite{Grbic05}. Thus the single-level regime becomes accessible for the first time.


The host heterostructure consists of a 5 nm undoped GaAs cap layer, followed by a 15 nm thick, homogeneously C-doped layer of AlGaAs, which is separated from the two-dimensional hole gas (2DHG) by a 25 nm thick, undoped AlGaAs spacer layer \cite{Reuter99}. It is to be noted that C acts as an acceptor on the (100) plane \cite{Wieck00}, and thus the anisotropy in the 2DHG formed in this plane is significantly reduced compared to the case of Si doped (311) heterostructures. The functionality of devices patterned on C-doped GaAs wafers as well as the interpretation of the transport experiments in these devices are therefore expected to be independent of the particular orientation of the device with respect to the wafer. Prior to sample fabrication the quality of the 2DHG was characterized by standard magnetotransport measurements at 4.2K and the following values were obtained for the hole density and mobility: $n$ = 4$\times$10$^{11}$ cm$^{-2}$, $\mu$ = 120'000 cm$^{2}$/Vs . The in-plane effective masses of the two spin-split subbands in similar C-doped GaAs heterostructures are $m_{1}$ = 0.34 $m_{e}$ and $m_{2}$ = 0.53 $m_{e}$ \cite{Grbic04}. Typical values for the interaction parameter are estimated to be $r_{s} \geq$ 5.

\begin{figure}[t!]
\includegraphics[width=\columnwidth]{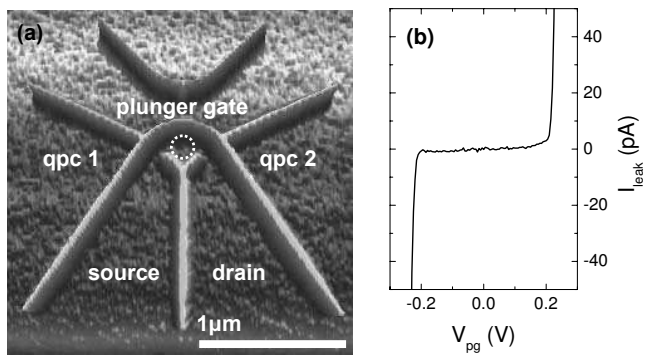}
\caption{(a) AFM image of the sample topology. The 15 nm high oxide lines are prepared by local anodic oxidation and create insulating barriers in the underlying 2DHG separating it into electrically disconnected areas. The dashed circle indicates the location of the QD which is connected to source and drain contacts by two QPCs. The couplings of the dot to the leads are tuned individually by the in-plane gates qpc1 and qpc2 while the electrochemical potential of the dot can be varied by applying a voltage on the nearby plunger gate. (b) Typical breakdown characteristics of the oxide lines at 60 mK base temperature.}
\label{structure.fig}
\end{figure}

The sample was patterned in the 2DHG by local anodic oxidation lithography \cite{Held98, Rokhinson02}. The bright oxide lines displayed in Fig.~\ref{structure.fig}(a) are created on the wafer surface by using the charged tip of an atomic force microscope (AFM) in a humid environment. Oxide lines as high as 15 nm are able to locally deplete the 2DHG situated 45nm below the surface separating the 2DHG into laterally disconnected regions which are individually connected to metallic leads. Voltages in the range of $[$-200 mV, +200 mV$]$ can be applied between separated regions without any significant leakage current across the oxide line, as illustrated in Fig.~\ref{structure.fig}(b). A small quantum dot is formed between the source ($S$) and drain ($D$) leads. The coupling of the dot to $S$ and $D$ through two quantum point contacts (QPCs) can be tuned individually by applying voltages on the nearby in-plane gates $qpc1$ and $qpc2$. The third in-plane plunger gate $PG$ is used to align the electrochemical potential and thus the number of the confined holes in the dot with respect to the electrochemical potentials in $S$ and $D$.\cite{footnote}

The transport experiments were carried out at 60 mK base temperature of a standard $^{3}$He/$^{4}$He dilution refrigerator. The two-terminal electrical conductance was measured by the simultaneous application of a symmetrical ac bias with an amplitude of 20 $\mu$V at 31 Hz lock-in frequency and dc biases up to 2 mV between $S$ and $D$. The resolution of the current detection was better than 20 fA at 0.5 Hz bandwidth.

\begin{figure}[t!]
\includegraphics[width=\columnwidth]{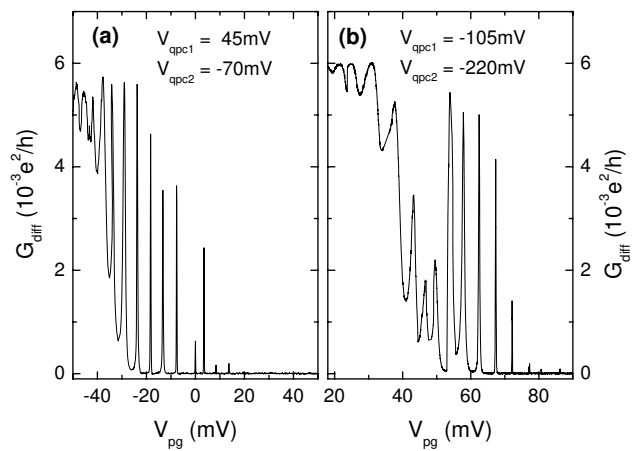}
\caption{Zero bias differential conductance through the dot measured at 60mK base temperature as a function of the plunger gate voltage at two different gate configurations (a) [$V_{\rm qpc1}$ = 45 mV, $V_{\rm qpc2}$ = -70 mV] and (b) [$V_{\rm qpc1}$ = -105 mV, $V_{\rm qpc2}$ = -220 mV], both in the symmetric coupling regime.}
\label{coulombpeaks.fig}
\end{figure}

From Coulomb diamond measurements we found that the dot is symmetrically coupled to the $S$ and $D$ leads when the control gates of the nominally 140 nm wide QPCs are tuned according to $V_{\rm qpc2}=V_{\rm qpc1}-115$ mV within the $[$-200 mV, +200 mV$]$ insulating regime of the oxide lines. In order to explore Coulomb blockaded transport the differential conductance of the dot was measured as a function of the $V_{\rm pg}$ plunger gate voltage at zero dc bias. The pronounced conductance resonances observed at different gate configurations are illustrated in Fig.~\ref{coulombpeaks.fig}. The fact that the dot closes with increasing $V_{\rm pg}$ confirms that the electrical transport is maintained by holes. At the two gate configurations selected in Fig.~\ref{coulombpeaks.fig} the peak positions were stable through several consecutive plunger gate sweeps on a time scale of a day with an accuracy of 0.1mV. However, at some particular gate configurations sudden rearrangements of the background charges made reproducible measurements difficult. From the relative shift of the resonances at different $V_{\rm qpc1}$ and $V_{\rm qpc2}$ voltages it is also clearly visible that the plunger gate not only tunes the number of holes in the dot but it also acts on the two QPCs leading to a change in the coupling strength of the dot to $S$ and $D$.

\begin{figure}[t!]
\center
\includegraphics[width=0.9\columnwidth]{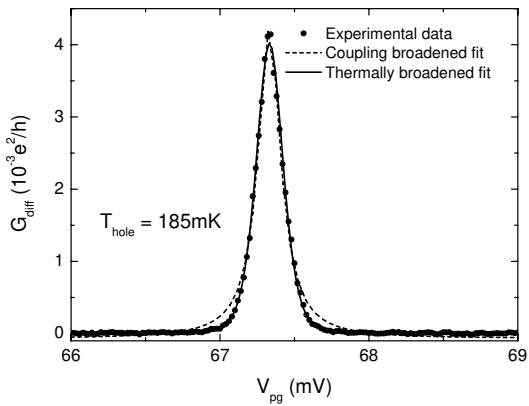}
\caption{Magnified view of the Coulomb peak at $V_{\rm pg}$ = 67.3 mV, $V_{\rm qpc1}$ = -105 mV, $V_{\rm qpc2}$ = -220 mV in Fig.~\ref{coulombpeaks.fig}(b). The experimental data (dots) are fitted to a thermally broadened (solid line) as well as to a coupling broadened resonance lineshape (dashed line). The thermally broadened model reveals a hole temperature of $T_{\rm hole}$ = 185 mK.}
\label{lineshape.fig}
\end{figure}

In the weak coupling regime each resonance line has been fitted with an expression for a thermally broadened Coulomb blockade peak in the multi-level as well as in the single-level transport regime according to \cite{Beenakker91}
\begin{equation}
G=G_{\rm max}\cdot\cosh^{-2}\left[\frac{\alpha e(V_{\rm pg}-V_{0})}{\nu k_{\rm B}T_{\rm hole}}\right] \mbox{,}
\label{cosh}
\end{equation}
where $G_{\rm max}$ and $V_{0}$ are the amplitude and the position of the Coulomb peak, respectively. The lever arm $\alpha$ of the plunger gate can be determined from Coulomb diamond measurements, as will be discussed later. The coefficient $\nu$ in the denominator of the $\cosh$ function equals 2 in the single-level regime while it is roughly 2.5 in case of multi-level transport \cite{Beenakker91}. For comparison a coupling broadened Lorentzian function was also fitted to each resonance \cite{Beenakker91}. The magnified view of a representative peak at $V_{\rm pg}$ = 67.3 mV in Fig.~\ref{coulombpeaks.fig}(b) as well as the fitted curves are shown in Fig.~\ref{lineshape.fig}. In all cases the thermally broadened resonance fits significantly better to the data than the coupling broadened resonance, confirming that the dot is indeed in the weak coupling regime and that the peak broadening is determined by temperature rather than by the coupling to the leads. From Coulomb diamond measurements (see below) we can also conclude that $k_{\rm B}T_{\rm hole}\approx\Delta\ll E_{\rm C}$, where $\Delta$ and $E_{\rm C}$ are the mean single-particle level spacing and the charging energy of the dot, respectively. This indicates that the dot is in the single-level transport regime and $\nu$ = 2 applies in Eq.~\ref{cosh}. As a fitting parameter we obtain typical hole temperatures in the range of T$_{\rm hole}=160-190$ mK. In case of symmetric coupling to the leads it is straightforward to evaluate the $\Gamma_{\rm S}\approx\Gamma_{\rm D}=\Gamma$ coupling strength from the $G_{\rm max}=e^{2}\Gamma/(8hk_{\rm B}T_{\rm hole})$ single-level transport formula for the resonance amplitude. Using the fitted values of $T_{\rm hole}$ we obtain $\Gamma\sim$ 1 $\mu$eV in agreement with the condition of the thermally broadened model.

\begin{figure}[t!]
\includegraphics[width=\columnwidth]{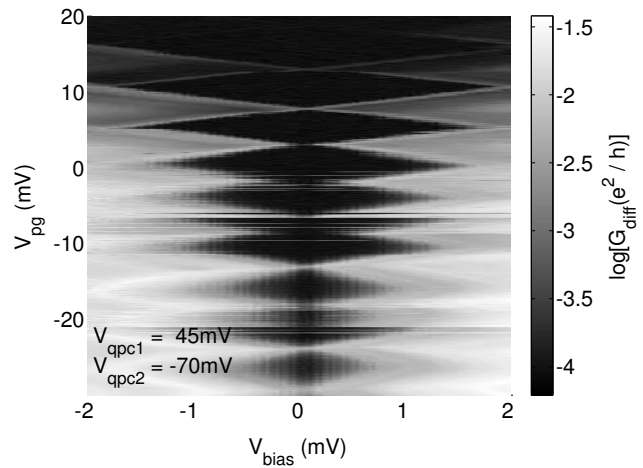}
\caption{Logarithmic gray scale plot of the Coulomb diamonds observed in the finite bias differential conductance of the dot (dark regions represent low conductance). The charging energy associated with the size of the diamonds is clearly enhanced at higher plunger gate voltages. The measurement was performed at $V_{\rm qpc1}$ = 45 mV and $V_{\rm qpc2}$ = -70 mV gate configuration at 60 mK base temperature.}
\label{diamonds.fig}
\end{figure}

Coulomb diamond measurements, i.e., measurements of the differential conductance as a function of bias voltage $V_{\rm bias}$ and plunger gate voltage $V_{\rm pg}$, were performed in the weak coupling regime. The result of an overall voltage scan is shown in Fig.~\ref{diamonds.fig}. The size of the diamonds clearly increases towards higher $V_{\rm pg}$ values which indicates the reduction of the electrostatic size of the dot, a phenomenon, that is characteristic to quantum dots with a small number of confined particles. From the extent of the diamonds in bias direction measured at the highest accessible $V_{\rm pg}$ regime we estimate $E_{C}\approx2$meV for the charging energy, while the lever-arm of the plunger gate is $\alpha\approx0.28$. This charging energy corresponds to a capacitance of the dot $C=e^2/E_{\rm C}\approx8\times10^{-17}$F. Attributing a disk-like shape to the dot, the capacitance is given by $C=8\varepsilon_{0}\varepsilon_{r}r$, where $r$ is the radius of the dot and $\varepsilon_{r}$ = 12.9 for GaAs . This enables the rough estimation of the electronic diameter of the dot to be $\approx$ 170 nm, which is in good agreement with the lithographic dimensions of the sample and indicates an upper limit of $\approx$ 90 for the number of the holes stored in the dot.

\begin{figure}[t!]
\includegraphics[width=\columnwidth]{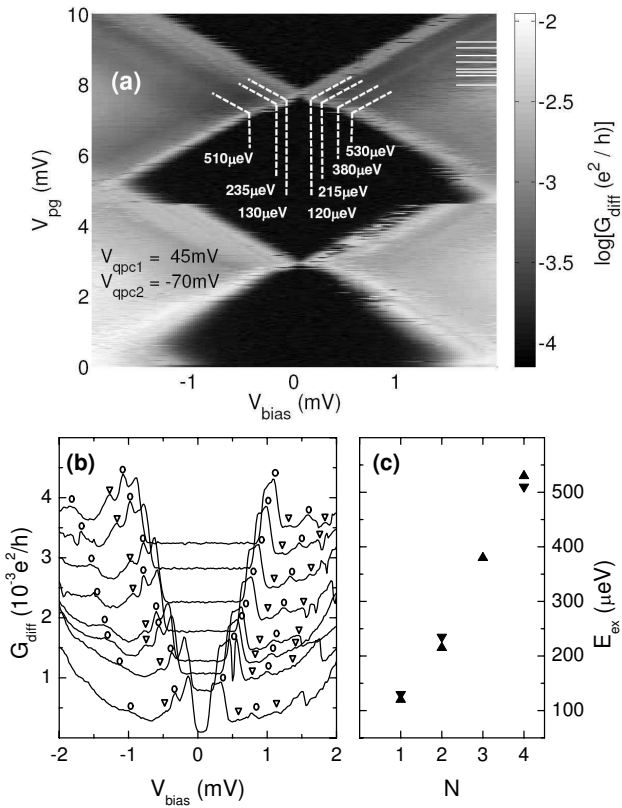}
\caption{(a) High resolution scan of the Coulomb diamonds displayed in Fig.~\ref{diamonds.fig} in the 0 $<$ $V_{pg}$ $<$ 10 mV regime. The parallel lines of higher conductivity outside the diamonds are attributed to sequential elastic tunneling through single-particle excited states of the dot. The corresponding excitation energies of the levels are also indicated. (b) Individual differential conductance traces recorded at various plunger gate voltages between $V_{\rm pg}$ = 8 and 10 mV indicated by the horizontal lines on the right hand side of panel (a). Each curve is vertically shifted for better visibility by an offset which is proportional to the corresponding $V_{\rm pg}$ value. The open circles and triangles indicate the local maxima which line up with the different excited state lines highlighted in the upper panel. (c) Excited state energies deduced from the lines at the right ($\blacktriangle$) and left ($\blacktriangledown$) sides in (a).}
\label{excited.fig}
\end{figure}

It is also visible in Fig.~\ref{diamonds.fig} that in certain dot configurations like at $V_{\rm pg}$ = -7 mV and -21 mV, low-frequency switching noise due to charge rearrangements in the sample becomes quite expressed. Since these switching events may occur on a time scale of one day, a reasonable resolution of the consequent Coulomb diamond measurements is determined by the size of the scanned area in the $[V_{\rm bias},V_{\rm pg}]$ parameter space. Therefore, in order to resolve conductivity peaks outside the Coulomb diamonds which are related to excited states we focused the high resolution conductance scans to a smaller regime in Fig.~\ref{diamonds.fig}. In case of a quantum dot with steep potential walls, the mean single-particle level spacing can be calculated as $\Delta=2\pi\hbar^{2}/gm^{*}A$, where $g$ is the degeneracy of hole states and $A$ is the electronic area of the dot. Taking into account the electrostatic size of the dot as deduced from the charging energy and assuming an effective mass of 0.53 $m_{e}$, the mean single-particle level spacing is estimated to be $\triangle\approx$ 20 $\mu$eV. Due to the large effective mass of holes this is one order of magnitude smaller than typical values in electron quantum dots, but still comparable to $k_{\rm B}T_{\rm hole}$. An accordingly refined Coulomb diamond scan was performed at the same gate configuration as for the data presented in Fig.~\ref{diamonds.fig}, and the result is displayed in Fig.~\ref{excited.fig}(a). Apart from a small displacement due to a background charge rearrangement at $V_{\rm pg}\approx$ 4.5meV the three diamonds in Fig.~\ref{excited.fig}(a) can be directly mapped to Fig.~\ref{diamonds.fig}. Outside the diamonds, parallel to the edges lines of higher differential conductance are visible. The presence of the lines is further visualized in Fig.~\ref{excited.fig}(b) by displaying various cross sections of the uppermost diamond along the $V_{\rm bias}$ axis. Apart from the missing left side counterpart of the right hand side line at $\approx$ 380 $\mu$eV which reveals the weak coupling of the excited states to the leads, the well visible left-right symmetry of the line distribution excludes an alternative explanation in terms of resonances in the random potential landscape of the leads. One can, therefore, unambiguously attribute these lines to elastic sequential tunneling through the single-level excited states of the dot. The weak coupling of the excited states made the simultaneous detection of second order current steps in the Coulomb blockaded region, characteristic to inelastic co-tunneling via the excited states difficult. However, a faint contrast in the middle of the Coulomb diamonds was found in the strongly coupled regime where the connecting excited state lines outside the diamonds were no longer resolvable.

While the broadening of the observed lines clearly exceeds both $k_{\rm B}T_{\rm hole}$ and the estimated $\Delta$ = 20$ \mu$eV, the energy of the lowest well distinguishable excited state is found to be 120$\pm$10 $\mu$eV. According to the above estimation this corresponds to an electronic dot diameter of $\approx$ 50 nm which implies an electronic area that is smaller by one order of magnitude compared to the value deduced from the charging energy and indicates a hole occupation number of $N\leq$ 10. Beside the inaccuracy of the applied simple models this discrepancy may reflect the general breakdown of the constant interaction model as well as the pronounced role of the strong hole-hole interactions at small occupation numbers. It is also to be noted that the value of the effective mass used for the estimation of the single-level spacing is taken from measurements of the extended 2DHG. It is conceivable that the effective mass of the carriers confined in the dot is different. Due to their large effective mass compared to electrons, valence band holes can screen the confining potential more efficiently which results in an effective hard wall potential. Depending on the dot geometry this leads to a single-particle level spacing that generally increases with $\propto N^{2}$. In contrast, the observed equidistant distribution of the excited state levels displayed Fig.~\ref{excited.fig}(c) recalls a parabolic two-dimensional confining potential which reveals weak screening, providing further experimental evidence for the small number of holes stored in the dot. In order to be able to determine the exact number of holes in the dot and thus explore the effects of hole-hole correlations experimentally, we plan to implement additional low noise QPC charge detection measurements in the close vicinity of the dot and work towards smaller dot sizes.


In conclusion, we fabricated a quantum dot on a p-type GaAs/AlGaAs heterostructure by AFM oxidation lithography. Clear and reproducible Coulomb resonances were observed at weak couplings to the leads. From the Coulomb diamond measurements a charging energy with a magnitude up to $\sim$ 2 meV was found at elevated plunger gate voltages indicating a small number of the confined holes. Lines of higher conductivity in the charge stability diagram outside the Coulomb blockaded region of the dot are resolved for the first time. They are attributed to sequential tunneling through single-hole excited states which may open new routes to the electrical manipulation of the individual hole spins.

\acknowledgments
This research was supported by the Swiss National Science Foundation, the German Science Foundation and the German Ministry for Science and Education. Two of us (A. D. W. and D. R.) thank the SFB491 and the BMBF nanoQUIT for financial support.


\begin{references}
\bibitem{footnote} It is to be noted that $PG$ can serve also as a non-destructive detector for sensing the charge state of the dot. Results of such charge read-out experiments will be reported elsewhere.
\end{references}
\end{document}